\documentstyle[aps,twocolumn,graphicx]{revtex}

\begin{document}


\title{A simple dynamical model with history dependence for a sand-pile experiment}

\author{Shuji Ishihara\thanks{shuji@complex.c.u-tokyo .ac.jp} and Kunihiko Kaneko\\          
{\small \sl Department of Pure and Applied Sciences}                                        
{\small \sl University of Tokyo, Komaba, Meguro-ku, Tokyo 153, JAPAN}\\}                    
\date{\today}


\maketitle
  
\begin{abstract}
A lattice dynamics model is proposed for the history dependence
observed in sandpile experiments.  The dependence of the stress distribution
on the preparation of the sandpile is explained as a dependence of certain
attractors on the preparation of the system.  
The model has three phases, but the history dependence
is shown to exist only in the phase where
a perturbation is amplified selectively rather than globally 
when propagating in the downflow direction.  
The condition for this history dependence
is given in terms of the spatial Lyapunov exponent.
\end{abstract}

~\\
PACS number(s): 45.70.Cc, 05.45.-a, 05.45.Ra

~\\
~\\

 
Physical systems exhibiting a ``history dependence'' in the sense that their state 
is not solely determined by existing environmental conditions
but dependent on how the system was externally driven in past are not uncommon.
Examples of such systems can be found  among
glasses, polymers, and granular materials, and of course, in biological systems.
Although hysteresis between two states is the simplest form of history dependence, 
here we examine cases where the evolution of a system may lead to many different states. 
Consequently, some variable(s) for the internal state can
be considered a type of `memory'. In a dynamical system, each memory state can be interpreted
as a different attractor. Hence it is necessary to study
how  each attractor is selected and how feasible it is for 
one attractor to switch to another attractor.
In order to do so, we take a spatially extended system, and discuss 
the macroscopic features of the history dependence.

A simple example of history dependence is given by
recent sand-pile experiments \cite{sandpile}.
In these experiments, it is found that the pressure profile formed by the stress chains in a sand pile 
strongly depend on how the sand beads are piled.
In the present Letter, we introduce a simple toy model inspired by the
experiment, and discuss some characteristic features of a system with history dependence.

In our toy model we take a mesoscopic approach by using a coarse-grained stress field
instead of taking a variable for each bead position.
A `stress variable' $x_{ij}$ is assigned to 
each site located on a two-dimensional triangular lattice $(i,j)$.
Note that due to gravity, the stress in a sand pile increases from upper to lower layers. 
Each site is associated with an intrinsic gravitational weight such that,
when excluding the stress from higher layers,
the stress variable relaxes according to the weight of the site
given by $\beta_{ij}$.
The stress of a layer is
transferred downwards by distributing the stress on each of its sites to 
the right and left neighbor sites in the next layer.
The amount of stress transferred from the upper sites that can be
supported by each of the
two lower sites is assumed to depend on the existing stress values of the lower sites.
The larger the existing stress value of a site is, the more effect the stress from the upper site will have.\\
Based on these arguments, we introduce the following model of coupled differential equations:

\begin{eqnarray}
\tau\frac{d x_{ij}{\scriptsize (t)}}{d t}&=&-x_{ij}{\scriptsize (t)}+\beta _{ij}+x_{l_ij\!-\!1}{\scriptsize (t)}\frac{[x_{ij}{\scriptsize (t)}]^{\alpha}}{[x_{ij}{\scriptsize (t)}]^{\alpha}+[x_{i\!-\!1j}{\scriptsize (t)}]^{\alpha}} \nonumber \\
&+&x_{l_i\!+\!1j\!-\!1}{\scriptsize (t)}\frac{[x_{ij}{\scriptsize (t)}]^{\alpha}}{[x_{ij}{\scriptsize (t)}]^{\alpha}+[x_{i\!+\!1j}{\scriptsize (t)}]^{\alpha}} \label{eqn:continuous} \\
 & \mathrm{with} &~~~l_i = \left\{
\begin{array}{ccc}
 i & \mathrm{for} &  \mathrm{even}~~j \\
 i-1 & \mathrm{for} &  \mathrm{odd}~~ j 
\end{array}
\right. \nonumber
\end{eqnarray}

\noindent where $\alpha$ is a positive parameter that characterizes the tendency of
a site with a higher existing stress value to be affected more strongly by an upper layer. 
Here the site $(l_i,j-1)$ is the upper left of the site $(i,j)$ and the site $(l_i+1,j-1)$ is the upper right.   
The time scale $\tau$ can be set at 1, by rescaling the time.  We consider the
homogeneous case here with a constant $\beta_{ij}=\beta$ (which can be set to 1
by rescaling the variable $x$).  The lattice size, unless mentioned otherwise, is
chosen to be $N$ for the horizontal direction with periodic boundary conditions, 
and $M$ for the vertical direction (sites in both the horizontal and vertical directions are
numbered from $0$ and hence $j=M-1$ is the bottom layer\cite{q-model}).

In our model the `weight' of the site $(i,j-1)$ is supported by the two neighboring 
sites in the $j$th layer, with  the ratio of $x_{ij}^{\alpha}/(x_{ij}^{\alpha}+x_{i-1j}^{\alpha})$. 
Corresponding to the `conservation law' of  momentum flux, the relation 
\begin{math}
\frac{d \langle x_{j} \rangle _i}{dt}=\langle x_{j-1} \rangle _i -\langle x_{j} \rangle _i +\langle \beta _{j-1}\rangle _i
\end{math}
holds, where $\langle \> \rangle _i$ denotes the average over the horizontal
direction for the $j$th layer, $\langle x_{j} \rangle _i =\frac{1}{N}\sum _i x_{ij}$ 
and $\langle \beta _{j-1}\rangle _i =\frac{1}{N} \sum _{i} \beta _{ij-1}=1$. 
Since the equation is governed by the relaxation term, 
the stress variables $x_{ij}(t)$  are attracted to a fixed point solution $x^*_{ij}$. 
As will be shown, there is a huge number of stable
fixed point solutions for the steady state equation if $\alpha \gtrsim 0.7$.
For a fixed point solution, the conservation law  implies that
$\langle x^*_j \rangle _i = \langle x^*_{j-1} \rangle _i+\langle \beta _{j-1} \rangle _i=\cdots=j+1$.

Now we study the history dependence in our model by recalling the sandpile experiments.
For a sand pile, it is known\cite{dips} that when the sand is piled from a small hole,
the pressure distribution at the bottom has a minimum in the center, while it has a maximum at the center
when the sand is scattered during the piling.
With this experiment in mind,
we choose the following operations to prepare a `sand pile'.

In our model, we represent the process of piling as the
addition of a weight $\beta_{ij}=1$ at a site, while
the stress variable $x_{ij}$ then evolves from 0.
Before piling a bead at a site, 
$\beta_{ij}=0$ and $x_{ij}=0$ were assigned.  Then, in order to simulate a piling process from
bottom to top, at each step, we choose a horizontal site $i$ randomly from 
a Gaussian distribution with the standard deviation  $W \times(N/2)$
and took the largest value $j$ such that the site $(i,j)$ is not vacant.   If both the
neighboring lower sites are occupied,  the weight $\beta$ is added at that site.
If at least one of the sites of the neighboring lower sites is vacant,
this (meso-scopic) bead falls down to the lower vacant site.
By adding weights successively, the dynamics is iterated to lead to a fixed stress field.


The case with 
$W\ll 1$ corresponds to a piling process from a thin hole,
while on the other hand the case with
$W>1$  corresponds to a piling process where the beads are scattered.

\begin{figure}[hbtp]
\begin{center}
\includegraphics[width=7.0cm]{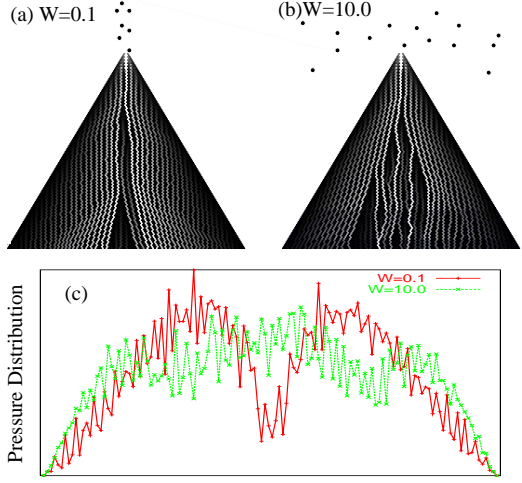}
\caption{Sand-pile simulations with (a)$W=0.1$(b)$W=10.0$. White pixels indicate large $x^{*}_{ij}$(i.e. $x^{*}_{ij} > \langle x^* \rangle _i$).
(c) Distribution of $x^*_{iM-1}$ obtained by averaging $100$ samples for $W=0.1$ (red line) and W=10.0(green line).
  The data are for $N=128$, but the results for $N=256$ are qualitatively identical after scaling.}
\label{fig:yama3}\end{center}
\end{figure}

Figure~\ref{fig:yama3} shows the results of the stress field
for (a)$W\!=\!0.1$, and (b)$W\!=\!10$. Note first, stress $x_{ij}$ is accumulated
on a few lattice sites forming a downflow stream. These sites
correspond to the stress chain observed in the sand pile experiments.
The accumulation of the stress itself is obvious, since our model has a tendency 
of more stress at a site with a larger stress value. In fact, the formation
of stress chains is commonly observed for $ \alpha \gtrsim 0.7$.

As shown in the figure the stress chain is distributed around the edge for the case (a)
(Figure~\ref{fig:yama3}(a)), while
in the case (b), it is distributed around the center.
The stress distribution at the bottom plate is plotted in Fig.\ref{fig:yama3}(c), 
obtained from the average of 100 samples.
It has a dip at the center for the case (a), while
it is unimodal  for (b).  
The stress distribution with the dip is observed independent of the system size and fits
a scale-invariant form .
These results agree well with the observations in sand-pile experiments \cite{sandpile}
and particle simulations.
Note that the history-dependence shown here
is generally observed  for $\alpha > 1$.

As a second demonstration of history dependence, we study how
the stress pattern depends on the time scale of preparing a sand pile.  
Differences in  time scales are introduced as follows: 
first the system is again prepared in  a ``no bead state '', i,e,, a state with
$\beta_{ij}=x_{ij}=0$ for all sites.  Then, every $T$ steps ( untill the  top layer $j=0$ is reached), 
we set $\beta=1$ for
all the sites in the lowest vacant layer simultaneously.
The stress field $x_{ij}$ of a new site is set to a random value in the range $[0,0.1]$
to represent small fluctuations.
Thus, with the help of eq.(\ref{eqn:continuous}), a fixed stress pattern $x^*_{ij}$ is obtained.

Through this process, stress chains are again formed.
We have computed the number of stress chains $\langle \langle  N_o \rangle \rangle$ by
counting the sites that satisfy  $x^*_{i j} > \langle x^*_{j} \rangle _i=j+1$ 
in the bottom layer, where
$\langle \langle ..\rangle \rangle$ is the ensemble average  over $500$ trials.  
When the piling process is fast ($T<10$), the number of stress chains is small, and in fact
the stress is accumulated at few sites in a lower layer.  For larger $T$ 
the number of stress chains is larger, and the weight is supported by more sites
\cite{com}.  This dependence on the time scale is commonly observed for $\alpha \!\!>\!\!1$.
The time scale  $T \approx 10$ is related to the
intrinsic time scale for the relaxation of the coupled system, as will be discussed later.
 
Now we discuss the origin of this history dependence in terms of dynamical systems theory.  
First, the coexistence of multiple attractors is confirmed by choosing a variety
of initial conditions with $x_{ij} \in [0,1]$.
Then after the relaxation is completed we compute the patterns of the fixed point 
attractors. 
Depending on the value of $\alpha$, we have the following three phases.

(I)$\alpha \lesssim 0.7$: The system
is always attracted to a stable homogeneous state. 
A single stable fixed-point with $x^*_{ij}=j+1$ (independent of $i$) exists.

\begin{figure}[tbph]
\begin{center}
\includegraphics[width=6.3cm]{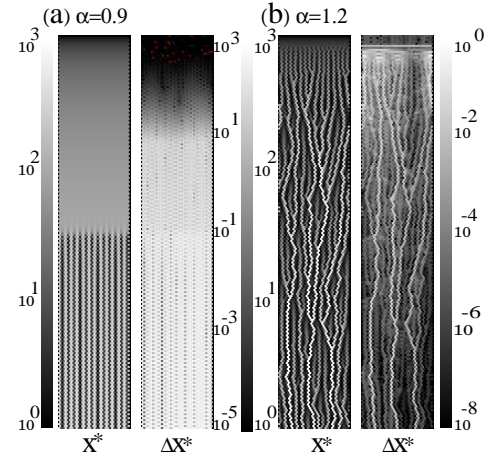}
\caption{The left columns in (a) and (b) show the stress variables $x^{*}_{ij}$ plotted in gray scale
for a system size of $32 \times 200$. (a)$\alpha =0.9$ and (b)$\alpha =1.2$.
The right columns depict $\Delta x^* _{ij}=| x_{ij}^{*'}-x^*_{ij}|$ in gray scale, where the $x_{ij}^{*'}$ is the fixed point reached after a
perturbation is added as $\vec{\beta} _0 \!=\! (\!1,\!1\!,\!\cdots\!)\!\to \!\vec{\beta} _0\! +\!\delta \vec{\beta} $, 
with $\delta \vec{\beta} \in$ [$-0.00001,0.00001$].}
\label{fig:push1layer}
\end{center}
\end{figure}

(II)$0.7 \lesssim \alpha \leq 1$:
Besides the homogeneous solution, many stable fixed point patterns appear (Fig.\ref{fig:push1layer}(a)).  
In the upper region, the patterns are almost homogeneous while in the lower region they are
striped (zigzag-like) with some dislocations.
For random initial conditions, the homogeneous state is not realized unless
we take an initial condition in its vicinity.  
 
(III)$\alpha \!>\!1$: The homogeneous solution becomes unstable at $\alpha =1+1/(j+1)$, and further downflow
it is always unstable.  When following the layers down, the behavior changes from an almost homogenous
state, to a stripe pattern as in phase II, and then to patterns where
stresses are accumulated into fewer sites forming inhomogeneous patterns
as shown in Fig.\ref{fig:push1layer}(b).  

Although both phases II and III have multiple fixed-points, 
the history dependence discussed so far is observed only for phase III.
One important  difference between phases II and III lies in
the response of each attractor to
a perturbation applied at the top layer. 

In phase II, there is a switch from one attractor to another when applying a tiny 
perturbation for at least $O(1)$ time units. A local perturbation 
in an upper layer spreads out horizontally in lower layers.
Fig.\ref{fig:push1layer}(a)  shows $x_{ij}^*$ and 
the variation $\Delta x_{ij}^*$ caused by the perturbation. 
As can be seen, $x_{ij}$ are altered for almost all sites.

In phase III, a perturbation needs to be applied for some time in order to
cause a switch to a different attractor. Furthermore, the length with which 
the perturbation needs to be applied
increases as the perturbation amplitude becomes smaller.  
Here, a perturbation is not
transferred smoothly to all sites, but is localized around
a few stress chains. 
As shown in Fig.\ref{fig:push1layer}, the sites whose  $x_{ij}$
values are altered are located around the stress chains, especially in the case of lower layers.   
The change of $x_{ij}^*$ is temporally intermittent.  After a perturbation is applied,
the stress values stay almost constant but then for some sites they show a
sudden increase, corresponding to a rearrangement of the stress chains. 

\begin{figure}[hbtp]
\begin{center}
\includegraphics[width=7.5cm]{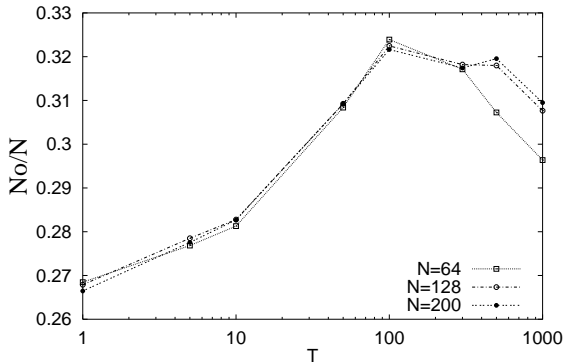}
\caption{The ratio of the number of stress chains and $N$ plotted against the time scale $T$ of the piling process described in the text.The vertical lattice size is $120$, and the horizontal size is 64($\Box$),128($\circ$),200($\bullet$). The results were obtained by averaging 500 runs. }
\label{fig:tumu} \end{center}
\end{figure}

\begin{figure}[tbhp]
\begin{center}
\includegraphics[width=7.5cm,height=4.5cm]{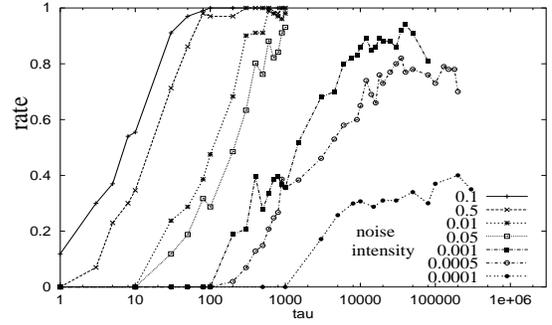}      
\caption{The fraction of attractor switches due to a perturbation with an amplitude $\epsilon \in$ [$0.0001$,$0.1$]. 
and a time duration $\tau$ in the top layer. The lattice size is $32 \times 200$ and $\alpha=1.2$. }
\label{fig:RMS.Dat}
\end{center}
\end{figure}

In order to study the relationship between the response to a perturbation
and the history dependence, we 
carried out the following simulations:
After the  system reaches a fixed point,
we apply a perturbation at the top ($0$-th) 
layer to change $\vec{\beta}_0=(1,1,\cdots)$ to 
$\vec{\beta}_0 +\delta \vec{\beta}_0 $ over $\tau$ steps, where
$\delta \vec{\beta}_0 $ is chosen randomly over $[-\epsilon,\epsilon]$ and fixed.
After waiting for the system to settle down to a fixed point attractor, we checked whether
it is identical to the original attractor or not.
By taking 100 samples,
the ratio of attractor switching is computed as a function of
the perturbation  strength $\epsilon$ and the perturbation duration $\tau$.
This is shown in Fig.\ref{fig:RMS.Dat} for phase III.
As the duration time $\tau$ for the application of the perturbation 
increases, its effect
accumulates  causing a switch from one attractor to another.
The required time duration  for a given switch-ratio 
roughly scales as $1/\epsilon$.  Hence the history dependence in Fig.\ref{fig:tumu} is
a natural consequence of the above time scale dependence.
Note that for  phase II, a dependence on $\tau$ or $\epsilon$ is not observed
since almost all perturbations kick the system
to a different attractor.

The response of a system against tiny perturbations is generally studied by means
of a Lyapunov analysis.
Ordinary Lyapunov analysis, however, is not useful here,
since all the fixed-points studied so far 
are linearly stable.  Rather,
in an open-flow system with a unidirectional interaction in space, 
the co-moving Lyapunov exponent 
measuring how a perturbation is amplified along the spatial direction is
important\cite{co-moving}.

In order to discuss the stability along the spatial direction,
we first obtain the 
relationship between the  fixed points at the 
  ($j-1$)-th layer and  the
$j$-th layer 
$\vec{x^*}_{j}=\vec{f}(\vec{x^*}_{j-1},\vec{x^*}_{j})$ by equating the r.h.s. of eq.(\ref{eqn:continuous}) with 0.
This gives a spatial map from $\vec{x^*}_{j-1}$ to $\vec{x^*}_{j}$.  The amplification rate
of the perturbation $\delta x^*_{j-1}$ at the $(j-1)$- th layer is given by $\delta x^*_{j}=M_j \delta x^*_{j-1}$,
with $M_j$ the Jacobi matrix for the spatial map. 

\begin{figure}[htpb]
\begin{center}
\includegraphics[width=8.0cm]{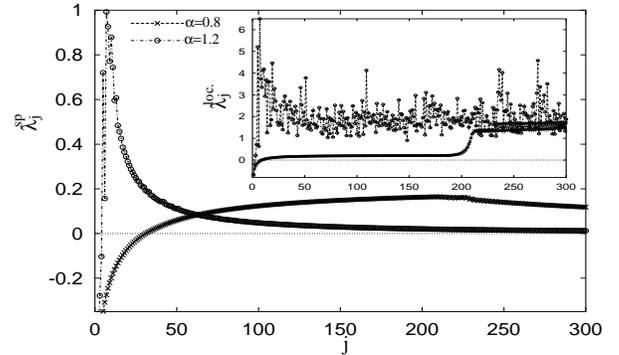}
\caption{Spatial Lyapunov exponent  $\lambda _{j}^{sp}$ and local spatial exponent (inset) for 
$\alpha=0.8$($\times$) and $1.2$($\circ$). The lattice size is $32 \times 200$.}
\label{fig:splyap}
\end{center}
\end{figure}

The rate at which a perturbation expands in the downflow direction is
then given by the maximal Lyapunov exponent of the spatial map
as $\lambda _{j}^{sp}\!=\! \log (\mu _j)/j $, 
where $\mu _j$ is square root of the maximal eigenvalue of 
$Q_j ^{\dagger}Q_j$, with $Q_j\!=\! M_j M_{j\!-\!1} \cdots M_1$i and $j$ starting from the $0$-th layer. 
The local expansion rate  from one layer to the next is
computed as the logarithm of the square root of the maximal eigenvalue of $M_j^{\dagger}M_j$.
We have plotted  $\lambda_j^{sp}$ in Fig.\ref{fig:splyap} as a function of the layer,
and $\lambda_j^{loc}$ in its inset.

In phases II and III ($\alpha\!\!=\!\!0.8,1.2$) $\lambda _{j}^{{\mathrm loc.}}$ 
is positive implying that a perturbation in one layer is amplified to the
next layer.
The  behavior of $\lambda _{j}^{sp}$, however, is different between the two phases.
In phase II (for $\alpha\!\!=\!\!0.8$)  $\lambda _{j}^{sp}$ increases with the layer $j$ 
and in fact the attractors of phase II are convectively unstable\cite{co-moving}.  
On the other hand, in phase III (for $\alpha\!\!=\!\!1.2$), 
$\lambda _{j}^{sp}$ first increases sharply  and then decreases 
towards $0$ as $j$ is increased.
Hence, a perturbation in the top layer may not reach
the bottom layer if the number of layers is sufficiently large.

This property is important for explaining the history dependence
shown in the piling process.  Since the local exponent is positive,
a perturbation  in the piling process is amplified to the next layer, and
thus leads to a variety of stress patterns. 
As the sand is piled, the distance between the top and
the bottom layers increases, and a perturbation by the addition of a layer at the top 
has less influence on the bottom layers because $\lambda _{j}^{sp}$ approaches zero.
The lower layers are thus `protected' from the
changes at the top layers.  Hence a given stress pattern selected in the
initial stage is `memorized'.  This leads to the history dependence 
in phase III, while for phase II, a perturbation in the top
layer is continuously amplified to the bottom 
(since $\lim_{j \rightarrow \infty}\lambda _{j}^{sp}>0$) and there is no history dependence.
 
In conclusion, we have constructed a simple model that describes the
history dependence of the pressure distribution in a sand-pile experiment. 
A condition for the history dependence is discussed from the viewpoint of 
a response against perturbations.  When there is a strong constraint on
the attractors that can be reached by a perturbation to a given  attractor, history 
dependence is observed, while if the
perturbation influences all sites globally, history dependence is
not possible \cite{attractor-switching}.  This condition for history dependence 
was further characterized by analyzing the spatial Lyapunov 
exponent.

The history dependence we have observed should be a general property of systems with 
the self-reinforcement of stress
(as represented by the $x^{\alpha}$ term in eq.(\ref{eqn:continuous}), but several other forms 
may also give this type of history dependence).  
How such a reinforcement term appears from a microscopic dynamics
of granular matter is left for future work, while it will also be important to generalize
the present study to other history dependent phenomena including  biological systems.

The authors would like to thank S. Sasa, K. Fujimoto, and A. Awazu for
discussions. We also thank F. H. Willeboordse for a critical reading of the manuscript. 
This research was supported by Grants-in-Aid for Scientific 
Research from the Ministry of Education, Culture, Sports,
Science and Technology of Japan (11CE2006).

\begin{thebibliography}{99}
\bibitem{com}
For larger $T$, it seems that the number of stress chains slightly decreases, whose
origin is not yet understood.
\end{thebibliography}
\end{document}